\documentclass[final,1p,times]{elsarticle} % do not change
\usepackage{amssymb}
\usepackage{graphicx}
\usepackage{lineno} % do not change

\newcommand{ \rts }{$\sqrt{s_{_{\rm NN}}}$}

\newcommand{\pttrig}{$p_{t}^{trig}$ }

\newcommand{\ptassno}{$p_{t}^{assoc}$}

\journal{Nuclear Physics A} % do not change
\begin{document} % do not change

\begin{frontmatter} % do not change

\title{Highlights from STAR (II)\\ - Hard Probes of the Initial and Final State -}
\author{J\"orn Putschke for the STAR Collaboration\footnote{For the full list of STAR authors and acknowledgements, see appendix in this volume.}}
\address{Yale University, Physics Department, New Haven, CT, USA, 06520-8120}

\ead{joern.putschke@yale.edu}

\begin{abstract}

Highlights of recent results from the STAR collaboration focusing on hard probes of the initial and final state are presented. New results at forward rapidities in d+Au collisions at low $x$ are utilized to study the possible onset of saturation effects at RHIC energies. New reference measurements, $\Upsilon$ production, nuclear-$k_T$ via di-jets in d+Au collisions, and jet quantities in p+p collisions will be discussed.  Final state nuclear modifications of $J/\psi$ and identified hadrons in heavy-ion collisions will be presented. In addition to di(multi)-hadron and direct $\gamma$-hadron correlations, new results from full-jet reconstruction in Au+Au collisions will be discussed with respect to the p+p reference measurements. These measurements can be used to put further constraints on the underlying mechanisms of partonic energy loss in heavy-ion collisions at RHIC.

\end{abstract}

\end{frontmatter} % do not change

%% QM09: we keep linenumbers at least for initial version
%\linenumbers % do not change

\section{Introduction}
\label{intro}

Hard probes - processes involving large momentum transfers - and their modification in Au+Au collisions with respect to p+p (d+Au) can be used to study the properties of the medium created in a heavy-ion collision at RHIC energies \cite{star_white}. In order to use hard processes as a well calibrated probe, one must be able to theoretically describe the reference measurements including initial and cold nuclear matter final state effects. The effect of parton energy loss in the final state is a well established observation in heavy-ion collisions at RHIC. It is evident in the suppression of the inclusive charged hadron and non-photonic electron yield at high-$p_T$, as well as in the associated yield in di-hadron correlations at high \pttrig and \ptassno, accompanied by an enhancement at low \ptassno, suggesting a dramatic softening of jet fragmentation \cite{star_highpTcoor,star2,npe}. However, these measurements are limited in their sensitivity due to well-known geometric biases (see for example \cite{Renk_Dihadron}). To overcome these biases one has to better constrain the parton kinematics, which is conceptually possible via direct $\gamma$- and full-jet reconstruction. New results utilizing these methods will be discussed.

In the following, new measurements addressing the initial state effects, such as the possible onset of saturation at low $x$, as well as improved reference measurements will be presented first. These will be followed by a discussion of how these calibrated hard probes become modified in the presence of  the dense matter created in heavy-ion collisions at RHIC.

\section{Hard Probes of the initial state and p+p reference measurements}
\label{initial}

\subsection{Forward rapidity measurements in p+p and d+Au collisions}

Forward rapidity measurements in d+Au collisions can be used to study the possible onset of saturation effects at RHIC energies.  The  Forward Meson Spectrometer (FMS), new in 2008 (p+p, d+Au), is an electromagnetic calorimeter that probes processes down to $x\approx10^{-4}$ at \rts=200 GeV at around $\eta\sim4$ \cite{ermes}, well into the range where saturation effects are expected to set in.
\begin{figure}[ht]
\begin{minipage}[t]{0.55 \textwidth}
\begin{center}
\includegraphics[width=1 \textwidth]{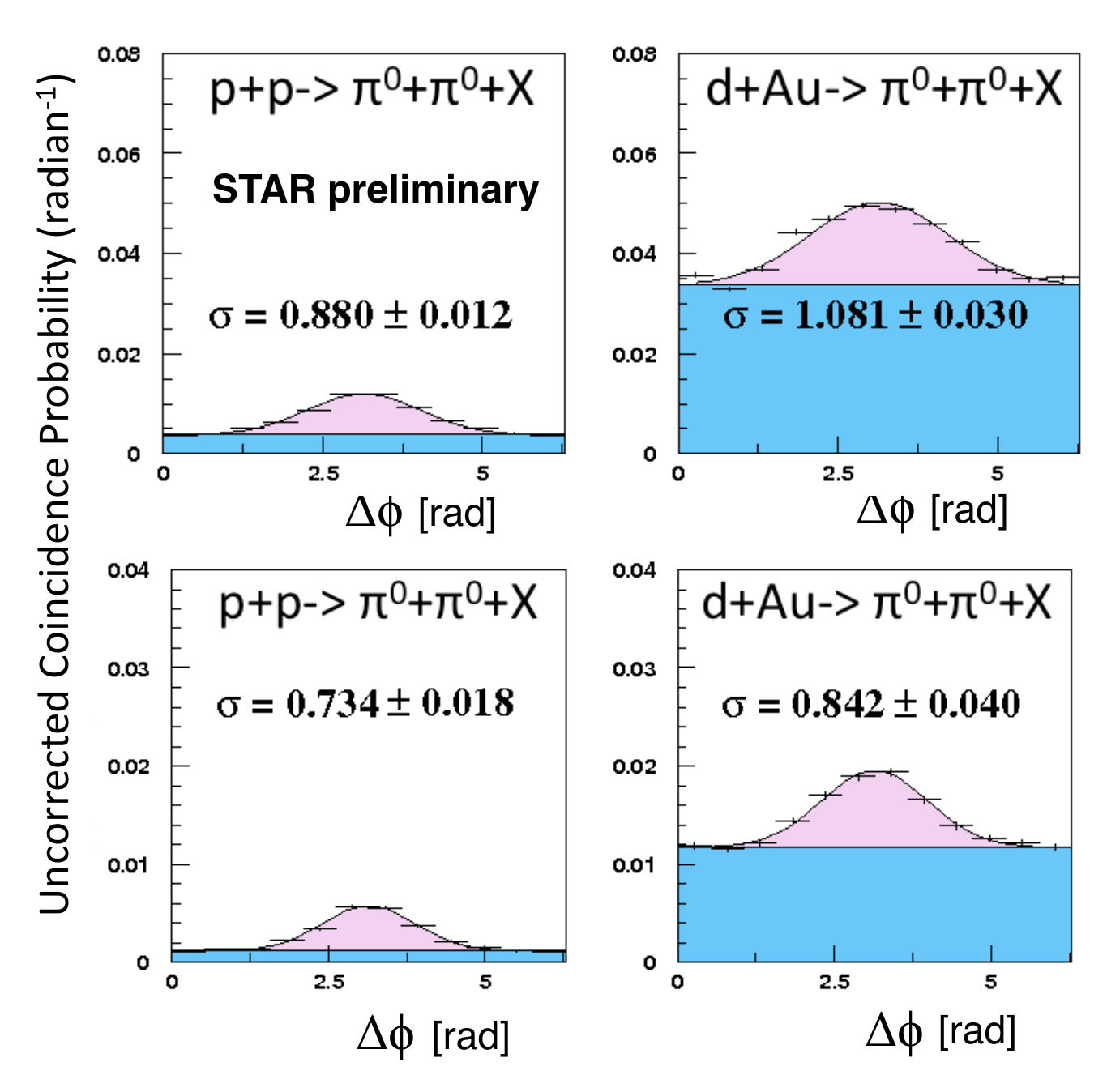}
\end{center}
\end{minipage}\hfill
\begin{minipage}[t]{0.4 \textwidth}
\vskip -6cm
\caption{\label{fms} (color online) Uncorrected coincidence probability versus azimuthal angle difference between a forward $\pi^0$ (measured in the FMS) and a mid-rapidity $\pi^0$ (measured in the EMC) in p+p and d+Au collisions at \rts=200 GeV. Upper row: $p_T^{FMS}<$ 2 GeV/c and  1 GeV/c $<p_T^{EMC}<p_T^{FMS}$; lower row: $p_T^{FMS}<$ 2.5 GeV/c and  1.5 GeV/c $<p_T^{EMC}<p_T^{FMS}$.}
\end{minipage} 
\vskip -0.35cm
\end{figure}

By measuring the azimuthal angular correlations, $\Delta\phi$, of a forward $\pi^0$ and a mid-rapdidty $\pi^0$ one expects a clear back-to-back peak in a 2$\rightarrow$2 parton scattering picture, whereas if saturation effects play a significant role a broadening of the recoil peak, or even its disappearance(monojet), could occur \cite{cgc}. These effects can be measured by comparing the $\Delta\phi$ correlations from p+p with d+Au collisions as shown in Fig.\ \ref{fms}. Two kinematic selections are shown in Fig.\ \ref{fms}, one with a more restrictive cut (bottom panel), suggested by pQCD calculations \cite{GSV}, the other with a lower $p_T$ cut closer to the saturation scale (top panel). The width of the recoil azimuthal correlation peak is larger for d+Au than for p+p and the difference ($\sigma_{dAu}-\sigma_{pp}$) increases from $0.11\pm0.04$ to $0.20\pm0.03$ for the lower $p_T$ cut, qualitatively consistent with a $p_T$ dependent picture of gluon saturation in the Au nucleus (more details in \cite{ermes}). In addition, the first observation of high-$x_F$ $J/\psi$'s  in p+p collisions at \rts=200 GeV have been reported by STAR at the conference \cite{perkins}.
 
\subsection{$\Upsilon$ measurement in d+Au collisions}

\begin{figure}[t]
\begin{minipage}[t]{0.4 \textwidth}
\begin{center}
\includegraphics[width=0.95 \textwidth]{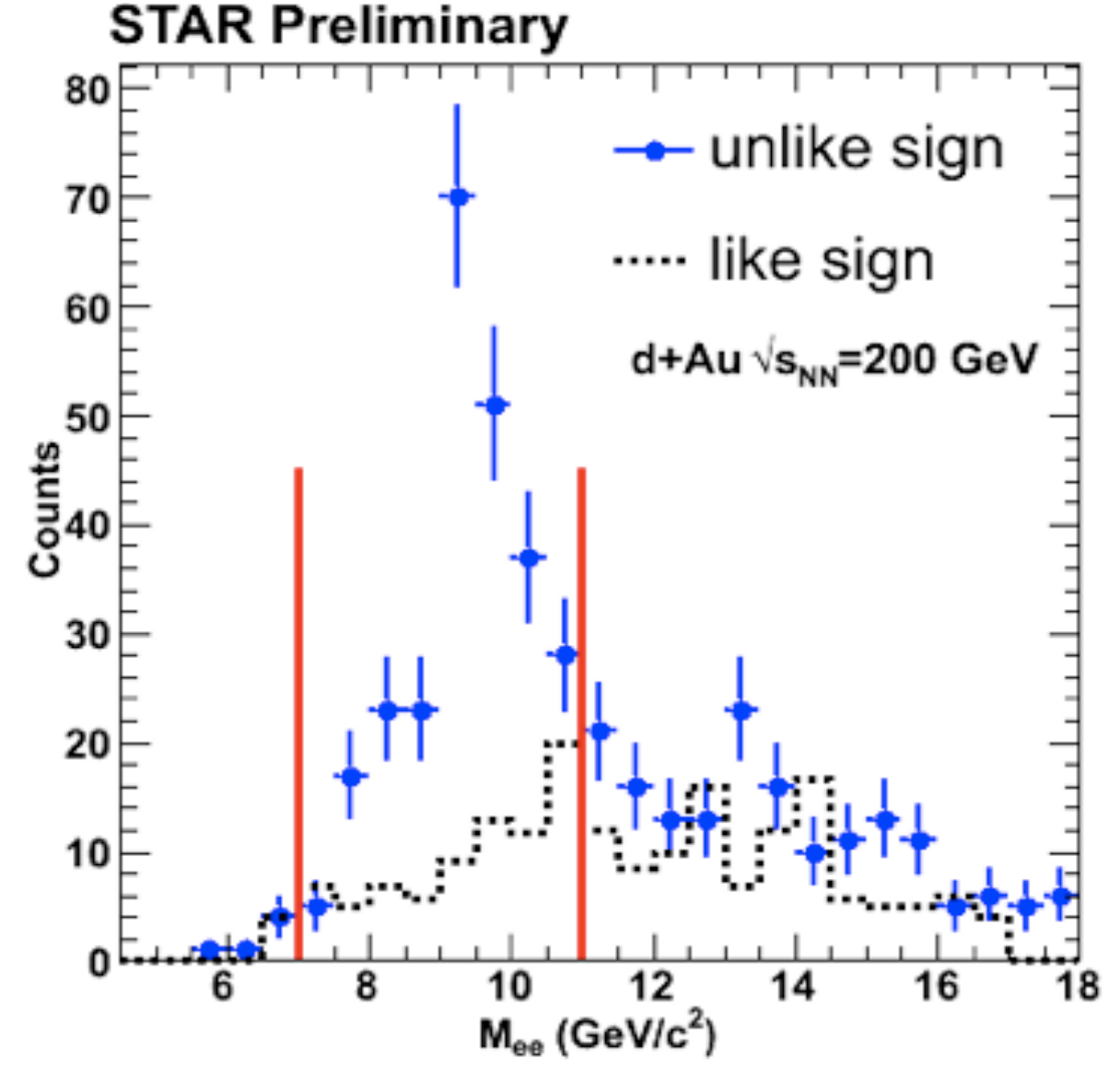}
\end{center}
\end{minipage}\hfill
\begin{minipage}[t]{0.5 \textwidth}
\vskip -4.5cm
\caption{\label{upsilon} (color online) $\Upsilon \rightarrow e^+e^-$ signal and background in \rts=200 GeV d+Au collisions. The solid symbols with statistical error bars are obtained by combining  the unlike-sign $ e^+e^-$ pairs. The dashed histogram shows the like-sign background.}
\end{minipage} 
\vskip -0.45cm
\end{figure}

The production of $\Upsilon$ and their modifications in heavy-ion collisions is a promising tool to study QGP properties at RHIC \cite{satz}. In addition to measurements in p+p and Au+Au it is important to study possible cold nuclear matter effects on  $\Upsilon$ production in d+Au collisions. The STAR $\Upsilon$ trigger enabled sampling a p+p equivalent luminosity of $\sim 12.5$pb$^{-1}$ for the 2008 d+Au run at \rts=200 GeV  \cite{haidong}. The measured signal of  $\Upsilon(1S+2S+3S)\rightarrow e^+ e^-$ shown in Fig.\ \ref{upsilon} has a significance of $\sim8\sigma$. The derived cross section at mid-rapidity is $B.R.\times d\sigma/dy=35\pm4(stat.)\pm5(sys.)$ nb, consistent with NLO calculations \cite{Frawley}. The nuclear modification factor $R_{dAu}=0.98\pm0.32(stat.)\pm0.28(sys.)$ suggests that $\Upsilon$ production follows binary scaling in d+Au collisions. With increased statistical significance these measurements will serve as an important reference for the upcoming measurement in Au+Au collisions (for more details see \cite{haidong}).

\subsection{Full-jet reconstruction in p+p and d+Au collisions}

The study of jet properties and the underlying event in p+p collisions is important for our understanding of QCD and the not well understood process of hadronization, as well as providing a baseline for jet measurements in heavy-ion collisions. 

\begin{figure}[ht]
\begin{center}
\includegraphics[width=0.51 \textwidth]{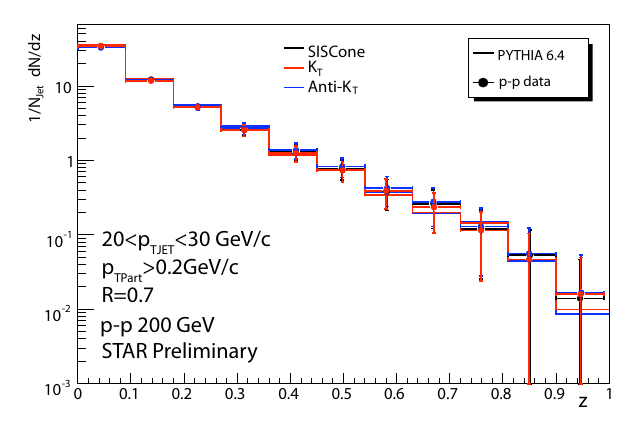}
\includegraphics[width=0.48 \textwidth]{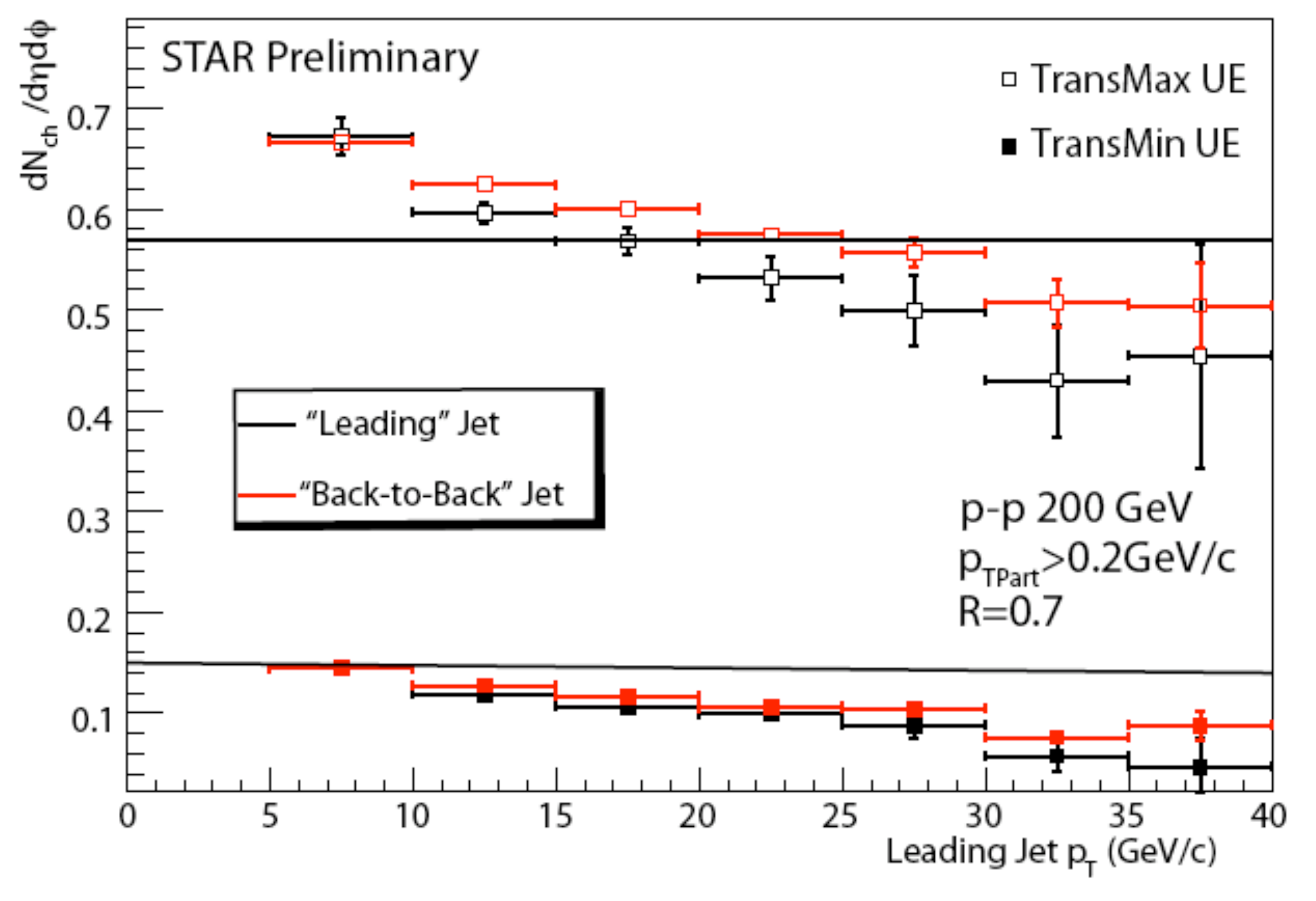}
\end{center}
\vskip -0.45cm
\caption{\label{ue}  (color online) Left panel: Charged particle, detector level, $z=p_T^{hadron}/p_{T,rec}^{jet}$ FF for jets reconstructed with 20 $<p_{T,rec}^{jet}<$ 30 GeV/c compared to PYTHIA for 3 different jet algorithms with $|\eta|<$1$-$R and R=0.7; right panel: the uncorrected charged particle density in the TransMin and TransMax regions as a function of reconstructed leading jet $p_{T}$, using the SISCone algorithm, with R=0.7.}
\end{figure}

Full jet reconstruction in STAR is performed by using charged particles, measured in the TPC, and neutral particles in the EMC at mid-rapidity. Modern jetfinding algorithms such as $k_T$ and Anti-$k_T$ recombination, and a seedless cone algorithm, SISCone, from the FastJet package \cite{fastjet} were used to reconstruct jets and to estimate systematic effects \cite{helen,ploskon,elena}. The jet energy resolution in p+p collisions can be obtained by either using PYTHIA simulations (PYTHIA $6.410$ Tune A \cite{pythia}) passed through STAR's simulation and reconstruction algorithms, or via the jet energy balance of back-to-back di-jets in p+p data. Both methods result in a comparable jet energy resolution of $\sim20$\% for reconstructed jet with $p_T>$10 GeV/c \cite{helen}. The uncorrected charged particle fragmentation function (FF) as a function of $z=p_T^{hadron}/p_T^{jet}$ for different resolution parameters, $R$, and jet energies show reasonable agreement with PYTHIA for all jet algorithms (see Fig.\ \ref{ue} left panel for $R=0.7$). This agreement especially for larger $R$ suggests that there are only minor NLO contributions beyond the PYTHIA LO calculations at RHIC energies. We also studied the underlying event (UE) in p+p collisions following the CDF technique \cite{UE}. Fig.\ \ref{ue} (right panel) shows the charged particle density in the UE. One can conclude that the UE is largely independent of the jet energy. By comparing different selections and subsets in the UE analysis, which have different sensitivities  to inital and final state radiation contributions, we observe only minor contributions from large angle intial/final state radiation (further details in \cite{helen}).

%d+Au without figure ...
Deploying full-jet reconstruction algorithms and methods to subtract background contributions in the FastJet package we performed a measurement of nuclear $k_{t}$ effects in d+Au collisions via di-jets \cite{jan}. To estimate the effect of jet energy resolution and possible background contributions of the underlying event in d+Au collisions on $k_{T,raw}=p_{T,1} \sin(\Delta\phi)$, we compared  PYTHIA simulations and PYTHIA jets (including detector simulation) embedded in d+Au minimum bias events.  This study suggests that jet energy resolution and possible background contributions in d+Au have only minor effects on the measured quantity \cite{jan}. We estimate a $\sigma_{k_{T,raw}}=3.0\pm0.1(stat)\pm0.4(sys)$ GeV/c in d+Au collisions at \rts=200 GeV. Further di-jet studies for different centralities in d+Au and p+p collisions will be pursued to estimate the nuclear $k_T$ effects.

% d+Au + figure ....
%Utilizing the tool of full-jet reconstruction and the methods to subtract background contributions in the FastJet package we performed a measurement to estimate the effect of nuclear $k_T$ in d+Au collisions via di-jets \cite{jan}. Fig.\ \ref{kt} shows the distribution of  $k_T=p_{T,1} \sin(\Delta\phi)$ for PYTHIA simulations (left panel) and PYTHIA jet embedded in a d+Au minimum bias event (middle panel), suggesting that the jet energy resolution and possible background contributions of the underlying event in d+Au collisions have only minor effects on the measured quantity. We estimate a $\sigma_{k_T}=3.0\pm0.1(stat)\pm0.4(sys)$ GeV/c in d+Au collisions at \rts=200 GeV (Fig.\ \ref{kt} right panel). Further di-jet studies for different centralities in d+Au and p+p collisions will be pursued to estimate the nuclear $k_T$ effects.

%\begin{figure}[t]
%\begin{center}
%\includegraphics[width=0.8 \textwidth]{dau_kt}
%\end{center}
%\vskip -0.45cm
%\caption{\label{kt} Distribution of $k_T=p_{T,1} \sin(\Delta\phi)$ for Anti-kt algorithm ($R$=0.5) using a HT4 trigger selection for  10 $<p_{T,2}<$ 20 GeV/c. Left panel: Pythia MC simulations; middle panel: Pythia+Geant embedded in a d+Au minbias event; right panel: d+Au measurements at \rts=200 GeV. }
%\end{figure}

\section{Hard Probes of the final state}
\label{final}

\subsection{Nuclear modification of $J/\psi$ at high-$p_T$ in Cu+Cu collisions}

Suppression of $J/\psi$ meson production in heavy-ion collisions has been proposed as a signature of QGP formation \cite{satz}. The measured suppression at \rts=200 GeV \cite{phenix} comparable to the observed magnitude at CERN-SPS energies is surprising and may be due to counterbalancing of larger dissociation with recombination of $c$ and $\bar{c}$ in the medium, which are more abundant at higher energies (see \cite{Frawley} and references within). To interpret the $J/\psi$ suppression, an understanding of quarkonium production in hadronic collisions is required. No model at present can fully explain the $J/\psi$ production in elementary collisions. $J/\psi$ measurements at high-$p_T$ in p+p and Au+Au collisions may provide further insight into the underlying process of quarkonium production \cite{perkins,jpsi}.
\begin{figure}[t]
\begin{minipage}[t]{0.55 \textwidth}
\begin{center}
\includegraphics[width=1 \textwidth]{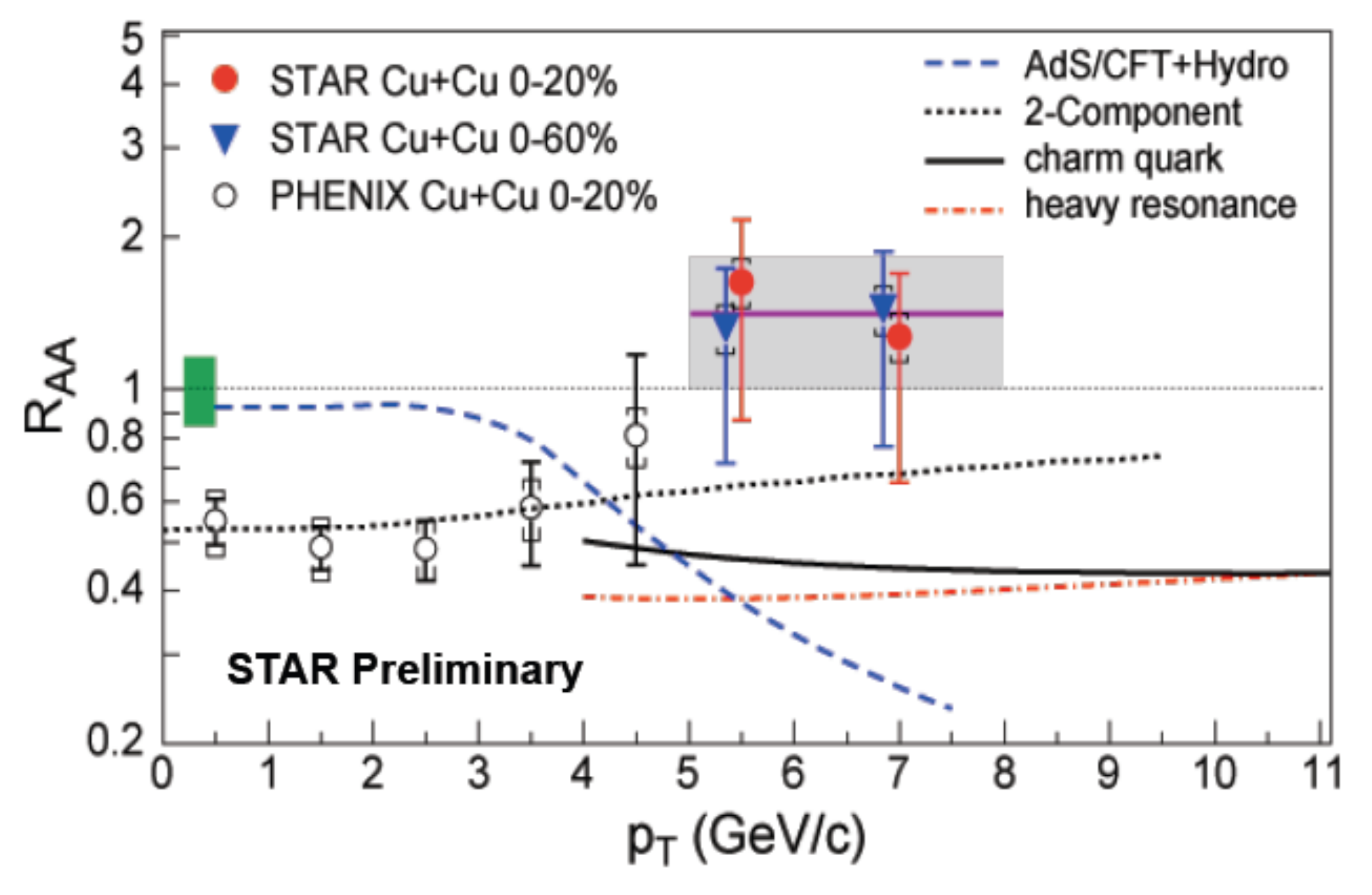}
\end{center}
\end{minipage}\hfill
\begin{minipage}[t]{0.4 \textwidth}
\vskip -4.75cm
\caption{\label{jpsi}  (color online) $J/\psi$ $R_{AA}$ vs.\ $p_T$ in Cu+Cu collisions at \rts=200 GeV \cite{highptjpsi}. STAR data points have statistical (bars) and systematic (caps) uncertainties. The box about unity on the left shows $R_{AA}$ normalization uncertainty, which is the quadrature sum of p+p normalization and binary collision scaling uncertainties. The solid line and band on the left plot show the average and uncertainty of the two 0-20\% data points.}
\end{minipage} 
\vskip -0.45cm
\end{figure}
In Fig.\ \ref{jpsi} the nuclear modification factor $R_{AA}$ for $J/\psi$ as a function of $p_T$ in Cu+Cu collisions at \rts=200 GeV is shown. The $R_{AA}$ increases as a function of $p_T$ and is consistent with unity for $p_T >$5 GeV/c. The $J/\psi$ is the only hadron measured in heavy-ion collisions at RHIC which does not exhibit a significant suppression at high $p_T$. The absence of $J/\psi$ suppression at high $p_T$, in contrast to a significant suppression of open charm, might indicate that high $p_T$  $J/\psi$ production is dominated by the color singlet channel (for more details on confronting theoretical calculations with the data see \cite{jpsi}). From high $p_T$  $J/\psi$-hadron azimuthal correlations in p+p collisions one can estimate the contribution from B-meson feed-down to inclusive $J/\psi$ production to be (13 $\pm$ 5)\% for $p_T >$5 GeV/c.

\subsection{Jet flavor conversion in Au+Au collisions}

To study the color charge effect of parton energy loss in heavy-ion collisions one can utilize the nuclear modification of identified hadrons ($\pi$, $K$ and $p$) at high $p_T$. 
\begin{figure}[t]
\begin{minipage}[t]{0.55 \textwidth}
\begin{center}
\includegraphics[width=1 \textwidth]{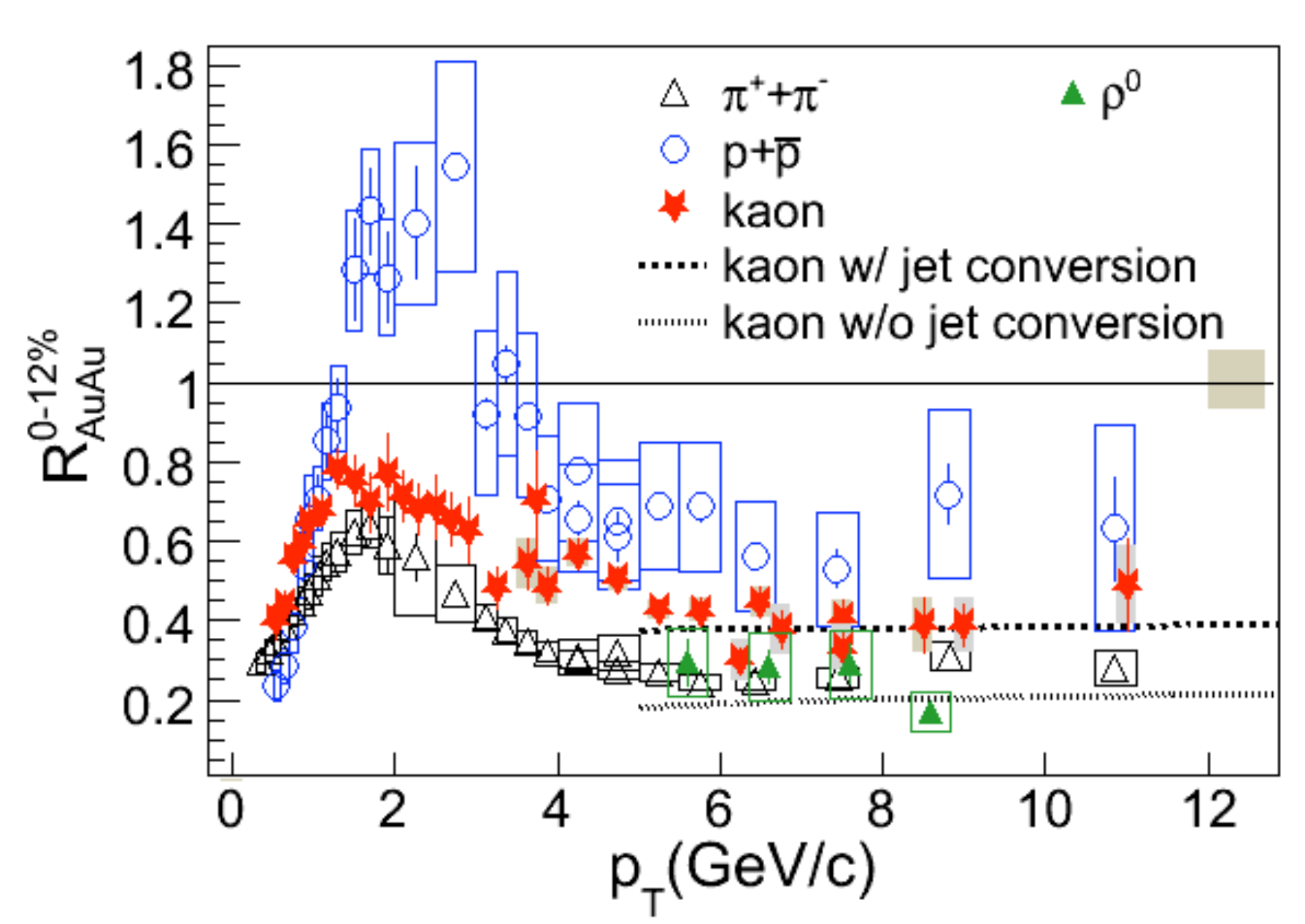}
\end{center}
\end{minipage}\hfill
\begin{minipage}[t]{0.4 \textwidth}
\vskip -5cm
\caption{\label{jet_flavor} (color online) Nuclear modification factors $R_{AA}$ of $\pi$, $K$, $p$ and $\rho$
in Au+Au collisions as a function of $p_T$. The bars and boxes represent statistical and
systematic uncertainties.}
\end{minipage} 
\vskip -0.25cm
\end{figure}
Figure\ \ref{jet_flavor} shows the nuclear modification factor $R_{AA}$ in central Au+Au collisions at \rts=200 GeV as a function of $p_T$ for $\pi$, $K$ and $p$ \cite{jet_flavor}. One observes that the $R_{AA}(K)$ and $R_{AA}(p)$ are larger than $R_{AA}(\pi)$ in contradiction to predictions from energy loss calculations \cite{wang}. The observed ordering of $R_{AA}$ for identified hadrons is consistent with predictions from calculations including jet flavor conversion in the hot dense medium \cite{fries} (see dashed lines in Fig.\ \ref{jet_flavor}). In addition the identified hadron high-$p_T$ spectra in p+p collisions are in agreement with NLO calculations and can be used to put constraints on the fragmentation functions (detailed discussion in \cite{jet_flavor}).

\subsection{Di- and Multi-hadron correlations}

\begin{figure}[t]
\begin{center}
\includegraphics[width=0.475 \textwidth]{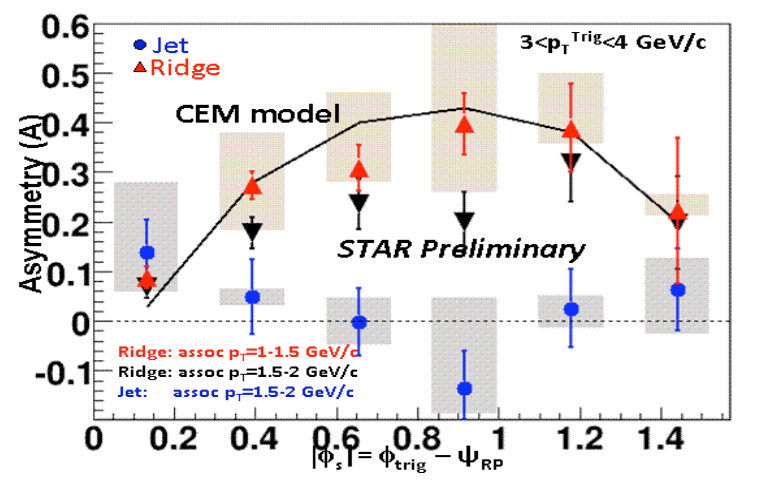}
\includegraphics[width=0.475 \textwidth]{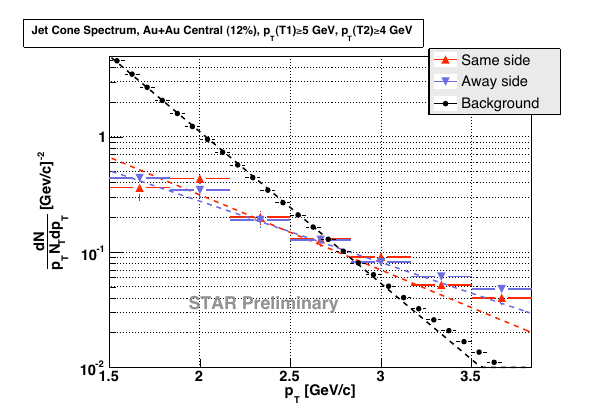}
\end{center}
\vskip -0.55cm
\caption{\label{ridge} (color online) left panel: Asymmetry parameter for jet and ridge as a function of $\phi_s$ for 20-60\% Au+Au
collisions. The shaded bands represent the systematic uncertainties in the flow estimation. Also shown as the black curve is the qualitative prediction from CEM (see text); right panel: Associated $p_T$  spectra for the symmetric di-jet triggered correlations: 5 $<p_{t}(T1)<$ 10 GeV/c and $p_{t}(T2)>$ 4 GeV/c with $p_T(A)>$ 1.5 GeV/c. The red triangles represent the near-side and the blue inverted triangles the away-side correlation. Minimum bias background in black solid circles.}
\vskip -0.35cm
\end{figure}

The observation of the near-side ridge (long range $\Delta\eta$ correlation) has motivated many theoretical attempts to explain this phenomena (for a summary see \cite{pawan}). A recent model, the Correlated Emission Model (CEM) \cite{rudi}, suggested that the ridge is formed from correlated particle emission due to aligned jet propagation and medium flow. As a consequence it predicts an asymmetric ridge $\Delta\phi$ correlation depending on the orientation with respect to the reaction plane $\phi_S=\phi_{trig}-\Psi_{RP}$. The measured ridge asymmetry (for details see \cite{pawan}) as shown in Fig.\ \ref{ridge} (left panel) is qualitatively consistent with the CEM predictions suggesting that the ridge formation might be caused by jet-flow alignment. Updates concerning 3-particle near-side $\Delta\eta$-$\Delta\eta$ correlations, using the charge properties of the associated hadrons to separate the ridge and jet components, can be found in \cite{pawan}. To study if the ridge is also present on the away-side, a technique was developed where one uses a pair of correlated high-$p_T$ hadron triggers to determine the jet-axis and study the associated hadron distribution (A) with respect to both triggers (T1 and T2) (details concerning the method can be found in \cite{kauder}). Selecting a ``symmetric" di-jet triggered case, 5 $<p_{t}(T1)<$ 10 GeV/c and $p_{t}(T2)>$ 4 GeV/c, with $p_T(A)>$ 1.5 GeV/c one observes no modification of the $\Delta\eta$ and $\Delta\phi$ correlations with respect to d+Au reference measurements, indicating that for this kinematical selection there is no evidence of a ridge or a conical structure on the away-side. For the symmetric di-jet selection the associated $p_T$ spectra for both trigger sides are consistent (Fig.\ \ref{ridge} right panel) and in agreement with d+Au measurements suggesting no, or only little, energy loss in the medium (more details in \cite{kauder}).

\subsection{Constraining the parton kinematics}

The measurements discussed above are limited in their sensitivity to jet quenching effects due to well-known geometric biases. To gain sensitivity and to be able to distinguish between different underlying energy loss mechanisms, one has to better constrain the parton kinematics. This can be achieved via direct $\gamma$- and full-jet reconstruction measurements.

\subsubsection{Direct $\gamma$-hadron fragmentation functions} 

\begin{figure}[t]
\begin{minipage}[t]{0.31 \textwidth}
\begin{center}
\includegraphics[width=1 \textwidth]{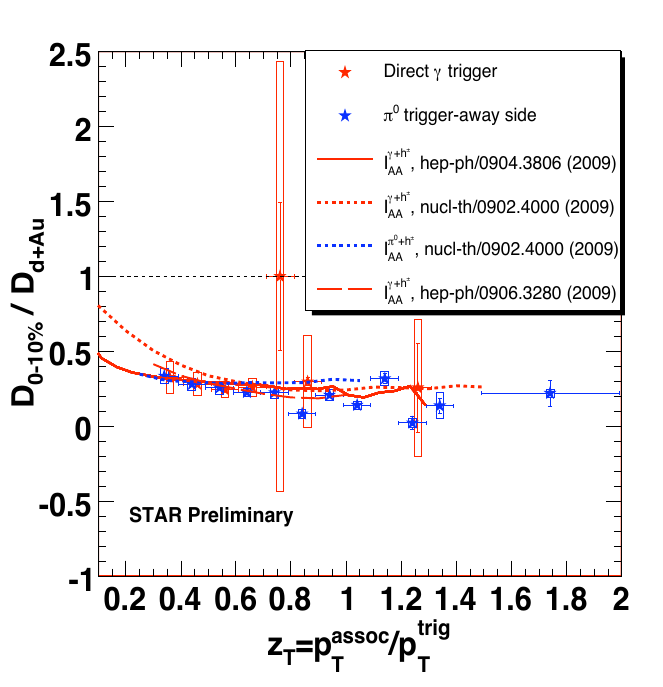}
\caption{\label{gamma}  (color online) Ratio $D_{0-10\%}/D_{d+Au}$ of $\pi^0$ and direct $\gamma$ (8 $< p_T^{\gamma,\pi^0} <$ 16 GeV/c) as function of $z_{T}$ at \rts=200 GeV compared to theoretical calculations (see text).}
\end{center}
\end{minipage}\hfill
\begin{minipage}[t]{0.31 \textwidth}
\includegraphics[width=1.1 \textwidth]{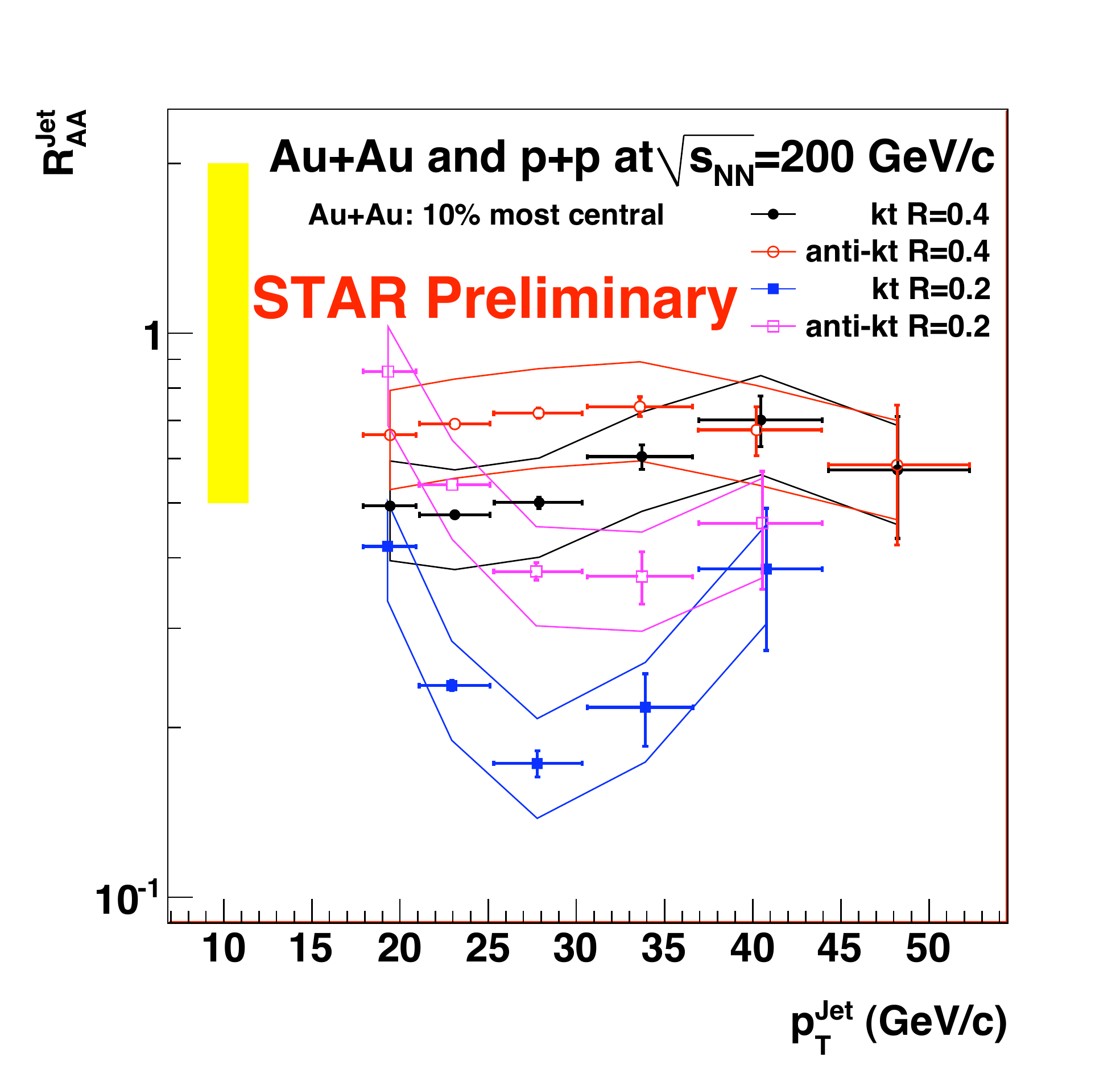}
\caption{\label{raa}  (color online) $R_{AA}$ of jets reconstructed with $k_T$ and anti-$k_T$ algorithms for the two different resolution parameters.}
\end{minipage} \hfill
\begin{minipage}[t]{0.31 \textwidth}
\includegraphics[width=1.1 \textwidth]{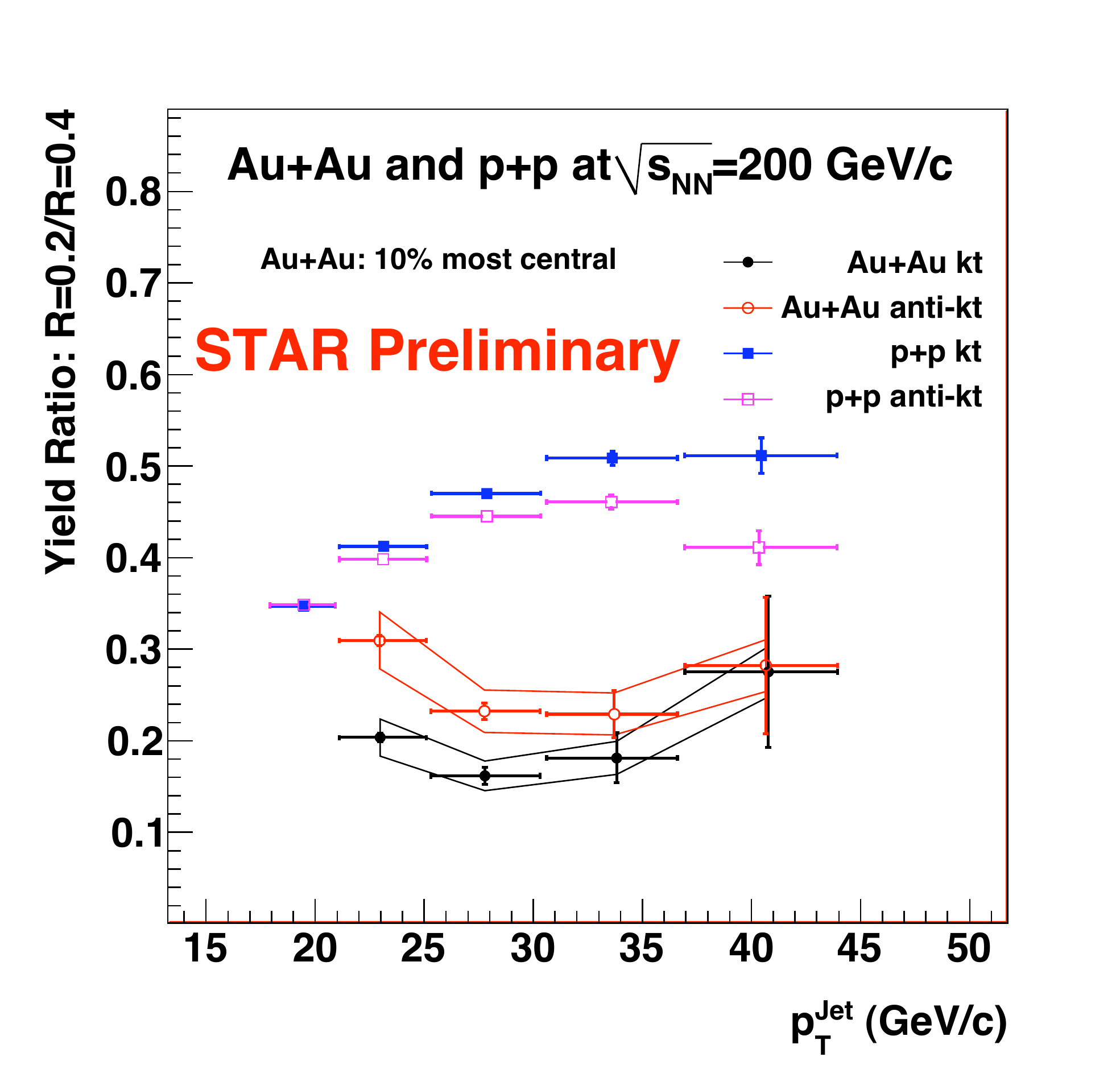}
\caption{\label{jet_shape}  (color online) Jet cross-section in p+p and Au+Au reconstructed with $R=0.2$ divided by cross-sections reconstructed with $R=0.4$ for $k_T$ and anti-$k_T$ algorithms .}
\end{minipage} \hfill
\vskip -0.475cm
\end{figure}

Conceptually clean measurements to constrain the parton kinematics are direct $\gamma$-hadron (jet) correlations. Studies are however, limited in statistics, due to the small cross-section of that process. Details concerning the analysis can be found in \cite{direct}. In Fig.\ \ref{gamma} the ratio of the integrated yield of associated charged hadron ($h^{\pm}$) per trigger in Au+Au relative to d+Au on the away-side ($D_{0-10\%}/D_{d+Au}$) for $\pi^0$ and direct $\gamma$ (8 $< p_T^{\gamma,\pi^0} <$ 16 GeV/c) as function of $z_{T}=p_T^{assoc}/p_T^{trig}$ at \rts=200 GeV is shown. In the measured $z_T$ kinematics there is no apparent difference in the suppression of $\pi^0$-$h^{\pm}$ and  direct $\gamma$-$h^{\pm}$ correlations. Naively one would expect a difference due to the trigger bias of the high-$p_T$ $\pi^0$'s, which would result in a longer pathlength of the recoil jet in the $\pi^0$-$h^{\pm}$ case. The absence of this effect might be due to a significant contribution of punch through or tangential jets in the measured kinematical regime, which would leave the fragmentation function unmodified. The measurements are in agreement with theoretical calculations (see \cite{direct} and references within), but to distinguish between the different models one has to extend the measurement to lower $z_T$, which we plan to pursue.

\subsubsection{Full-jet reconstruction in Au+Au collisions}

\begin{figure}[t]
\begin{center}
\includegraphics[width=0.45 \textwidth]{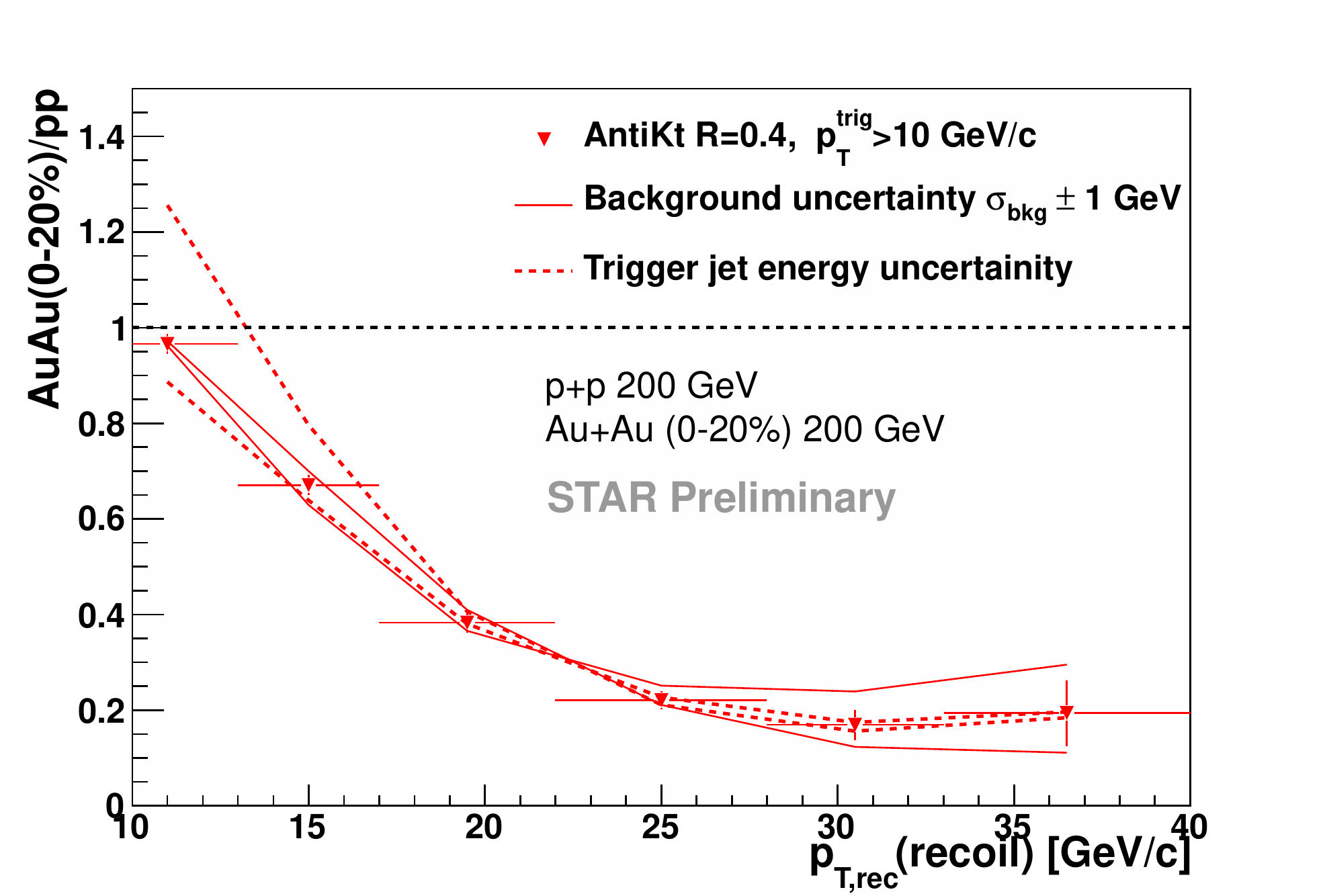}
\includegraphics[width=0.425 \textwidth]{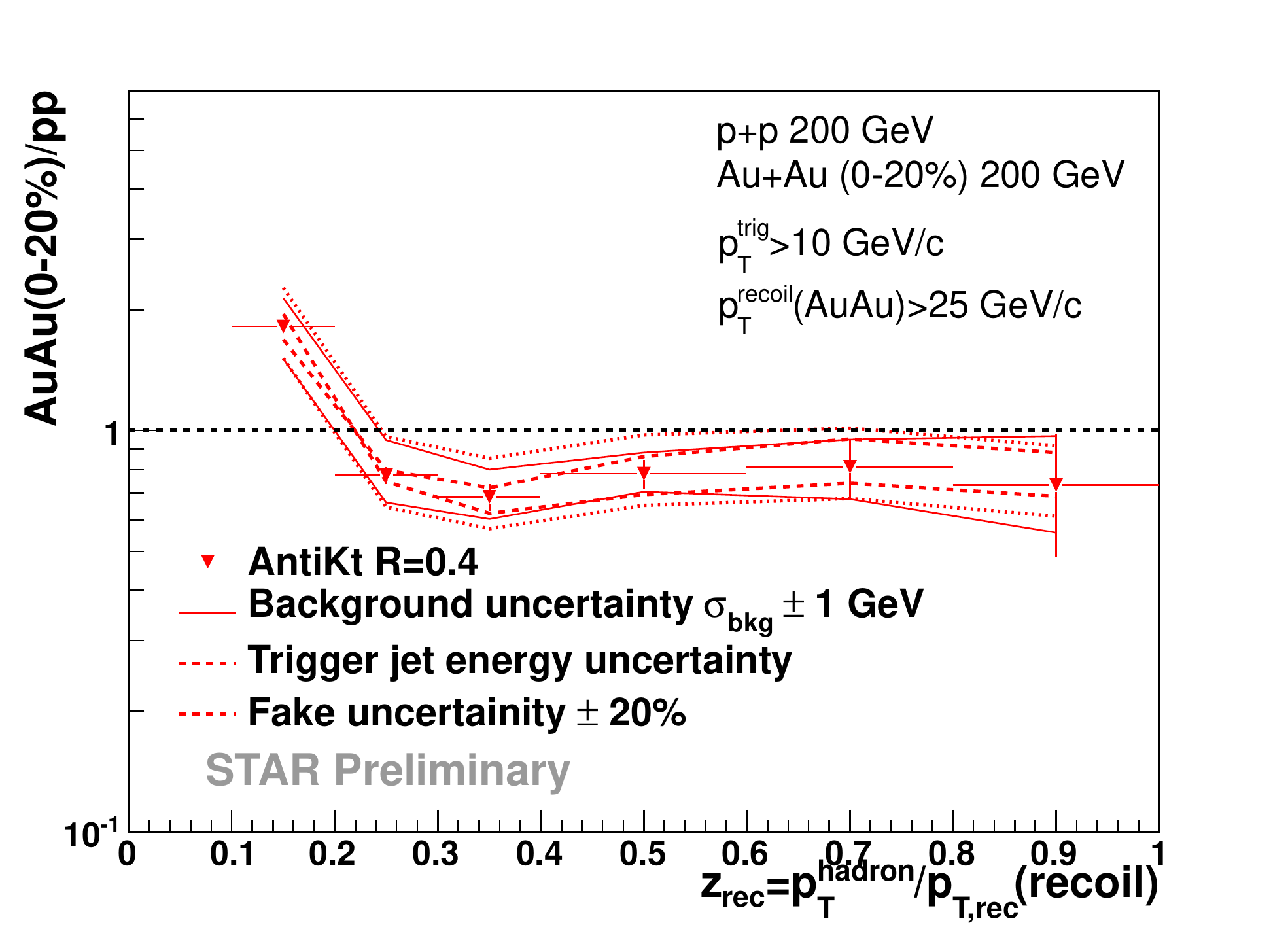}
\end{center}
\vskip -0.45cm
\caption{\label{ff} (color online) Left panel: Ratio of di-jet coincidence in Au+Au over p+p reference measurements ; right panel: Fragmentation function ratio Au+Au over p+p as a function of $z$ for $p_T^{recoil}>25$ GeV/c.}
\vskip -0.35cm
\end{figure}

Full-jet reconstruction in heavy-ion collisions is an alternative approach to constrain the parton kinematics over a larger range of jet energies as compared to direct $\gamma$-hadron (jet) correlations. The challenges in these measurements are to correct for contributions from the underlying heavy-ion event. In central Au+Au events at \rts=200 GeV  the background energy in a cone of $0.4$ is $\sim 45$ GeV/c and fluctuations in the background are of the order of $\sigma=6-7$ GeV/c  \cite{ploskon, elena}. A data driven correction scheme for background contributions and detector effects was developed, including a first attempt to estimate the corresponding systematic uncertainties \cite{helen, ploskon, elena}. In the case of unbiased jet reconstruction in heavy-ion collisions, one would expect a scaling of the inclusive jet cross-section with the number of binary collisions $N_{bin}$. The results shown in Fig.\ \ref{raa} and Fig.\ \ref{jet_shape} changed as compared to what was shown at the Quark Matter Conference, for further discussion see \cite{ploskon}. In Fig. \ref{raa} the $R_{AA}$ of jets reconstructed with the $k_T$ and anti-$k_T$ algorithm \cite{fastjet} for resolution parameters $R=0.2$ and $0.4$ is shown as a function of jet $p_T$. It is notable that with full-jet reconstruction the kinematic reach to study jet quenching effects is greatly extended, up to jet $p_T$'s of $\sim 45$ GeV/c in central Au+Au collisions. For $R=0.4$ one recovers a larger fraction of jets as compared to the inclusive hadron $R_{AA}$ and to $R=0.2$. This measurement indicates that an unbiased jet population is not recovered for $R=0.4$ in central Au+Au collisions, even though the systematic uncertainties are still quite large. An alternative explanation to an overall absorption, could be a significant jet broadening beyond $R=0.4$, which would lead to an underestimation of the reconstructed jet energy in Au+Au with respect to p+p at \rts=200 GeV. Fig.\ \ref{jet_shape} shows, by comparing the cross-section ratios in p+p and Au+Au of jets found with $R=0.2$ normalized to $R=0.4$, that there is a broadening in the jet structure. Assuming a continuous evolution, this suggests that there might be a significant contribution for $R>0.4$. This assumption is in agreement with the observed significant suppression in the di-jet coincidence measurements (see Fig.\ \ref{ff}, left panel),  where the path-length and therefore the effect of parton energy loss of the recoil jet is maximized by utilizing the trigger bias of the online triggered events (see \cite{elena}). The absence of strong modifications in the fragmentation function measurements (see Fig.\ref{ff}, right panel) can also be explained by a significant broadening in the jet structure due to energy loss. 

\section{Summary}
In these proceedings a wealth of new measurements from the STAR collaboration were presented utilizing hard probes to study the initial and final state effects. The possible onset of saturation effects, as well as new and improved baseline measurements of the initial state, were discussed. The presented data of nuclear modifications in the final state can be used to gain further insights into the production mechanisms of heavy-flavor. Full-jet reconstruction extends significantly the kinematical reach to study jet quenching effects and modifications of jet quantities in the presence of the hot, dense medium created in heavy-ion collisions at RHIC. These measurements will put further constraints on the underlying partonic energy loss mechanisms.

\bibliographystyle{epj}
\bibliography{ref}

\end{document}